\documentstyle[aas2pp4,psfig]{article}

\newcommand\mdot   {\hbox {${\dot M}$}}

\newcommand\pp     {$\pm$}
\newcommand\pers     {s$^{-1}$}
\newcommand\micros  {$\mu$s}

\righthead{Kilohertz Quasi-Periodic Oscillations in KS~1731$-$260}
\slugcomment{Submitted to ApJ Letters, February 1997}

\begin{document}

\title{Discovery of Two Simultaneous Kilohertz Quasi-Periodic
Oscillations in KS~1731$-$260}

\author{Rudy A. D. Wijnands and Michiel van der Klis} \affil{Astronomical
Institute ``Anton Pannekoek'', University of Amsterdam, \\ and Center
for High Energy Astrophysics, Kruislaan 403, NL-1098 SJ Amsterdam, The
Netherlands; rudy@astro.uva.nl, michiel@astro.uva.nl}

\begin{abstract}

We have discovered two simultaneous quasi-periodic oscillations (QPOs)
at 898.3\pp3.3 Hz and 1158.6\pp9.0 Hz in the 1996 August 1 observation
of the low-mass X-ray binary KS 1731$-$260 with the Rossi X-ray Timing
Explorer.  The rms amplitude and FWHM of the lower frequency QPO were
5.3\pp0.7~\% and 22\pp8 Hz, whereas those of the higher frequency QPO
were 5.2\pp1.0~\% and 37\pp21 Hz. At low inferred mass accretion rate
(\mdot) both QPOs are visible, at slightly higher \mdot~ the lower
frequency QPO disappears and the frequency of the higher frequency QPO
increases to $\sim$1178 Hz. At the highest inferred \mdot~ this QPO is
only marginally detectable (2.1 $\sigma$) near 1207 Hz, which is the
highest frequency so far observed in an X-ray binary. The frequency
difference (260.3\pp9.6 Hz) between the QPOs is equal to half the
frequency of the oscillations observed in a type I burst in this
source (at 523.92\pp0.05 Hz, Smith, Morgan and Bradt 1997). This
suggests that the neutron star spin frequency is 261.96 Hz (3.8 ms),
and that the lower frequency QPO is the beat between the higher
frequency QPO, which could be a preferred orbital frequency around the
neutron star, and the neutron star spin. During the 1996 August 31
observation we detected an additional QPO at 26.9\pp2.3 Hz, with a
FWHM and rms amplitude of 11\pp5 Hz and 3.4\pp0.6 \%.

\end{abstract}


\section{Introduction \label{intro}}

Kilohertz quasi-periodic oscillations (QPOs) have been found so far in
the persistent emission of nine low-mass X-ray binaries (LMXBs), three
Z sources (Sco X-1: van der Klis et al. 1996a, 1997b; GX 5$-$1: van
der Klis et al. 1996b; GX 17+2: van der Klis et al. 1997a), and six
atoll sources (4U 1636$-$53: Zhang et al. 1996, Wijnands et al. 1997;
4U~1728$-$34: Strohmayer et al. 1996; 4U~1608$-$52: Berger et
al. 1996; 4U~0614$+$09: Ford et al. 1996; 4U~1735$-$44: Wijnands et
al. 1996; 4U~1820$-$30: Smale, Zhang \& White 1996). In the sources
Sco X-1, GX 5$-$1, GX 17+2, 4U 1636$-$53, 4U 1728$-$34, 4U 0614$+$09,
and 4U 1820$-$30, two kilohertz QPOs are seen simultaneously, with a
frequency separation between 250 and 360 Hz; in 4U 1608$-$52 and 4U
1735$-$44 so far only one QPO has been observed. In the sources 4U
1636$-$53 and 4U 1728$-$34 (Zhang et al. 1997; Strohmayer et al. 1996)
oscillations were observed during type I bursts whose frequency (or
half of it) was consistent with being equal to the frequency
separation between the two kilohertz QPOs, which in some
beat-frequency models is interpreted as being the neutron star spin
frequency. Smith, Morgan and Bradt (1997) discovered coherent
523.92\pp0.05 Hz oscillations, which they interpreted as the neutron
star spin frequency, during a type I X-ray burst in the low-mass X-ray
binary KS 1731$-$260, and found no QPOs near 1 kilohertz, with upper
limits on the amplitude of 1~\%~rms. We reanalyzed part of the public
archive data of KS 1731$-$260, also used by Smith et al. (1997), and
report the discovery of two simultaneous QPOs in the persistent
emission near 898 Hz and 1159 Hz, with a frequency seperation
consistent with being equal to half the frequency of the oscillations
in the burst.

\begin{figure}[t]
\begin{center}
\begin{tabular}{c}
\psfig{figure=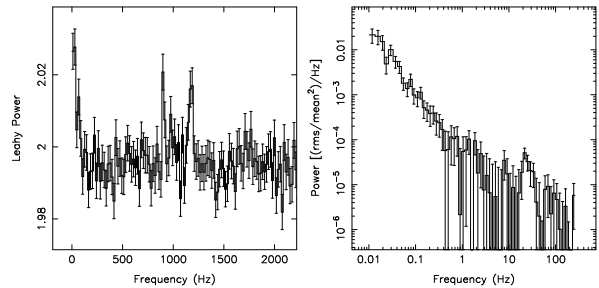,width=8cm}
\end{tabular}
\caption{Leahy-normalized power spectra (5.7--24.1
keV) of KS 1731$-$260 during the 1996 August 1 observation (left), and
the rms normalized power spectra (2.8--24.1 keV) during the 1996
August 31 observation (right). The left-hand figure is not corrected for
counting statistics or dead time.\label{powerspectrum}}
\end{center}
\end{figure}

\section{Observations and Analysis  \label{observations}}

We analyzed the public archive data of KS 1731$-$260 obtained by the
Rossi X-ray Timing Explorer (RXTE; Bradt, Rothschild \& Swank 1993) on
1996 August 1 1646--2044 UT and on 1996 August 31 1738 -- 1941 UT.
Due to Earth occultations and South Atlantic Anomaly passages the data
were split up in 3 segments during the August 1 observation, and 2
segments during the August 31 observation each with a duration of
2000--3000 s.  The (2.1--18.9 keV) source count rate varied between
1620 and 1990 counts \pers~ on August 1 and between 1970 and 2360
counts \pers~ on August 31. The background was typically 50 counts
\pers.  Fitting the spectra with a thermal bremsstrahlung model
(Sunyaev 1989, Smith et al. 1997), we derive the following 2-10 keV fluxes :
$3.2\times10^{-9}$ and $3.8\times10^{-9}$ ergs cm$^{-2}$ s$^{-1}$.

During both observations data were collected with a time resolution of
16 s (129 photon energy channels), and a time resolution of 62 \micros
~ (32 channels). Both modes covered the entire 2-60 keV range over
which RXTE's proportional counter array is sensitive. X-ray
color-color diagrams (CDs) were constructed from the 16 s data, and
power spectra were calculated from the 62 \micros ~ data using 16-s
and 256-s data intervals.  The power spectra were fitted using a power
law (the very low frequency noise [VLFN], measured between 0.01--1
Hz), Lorentzians (the QPOs, if present), a broken power law (the high
frequency noise [HFN], measure between 1--100 Hz), and a constant
level (the Poisson noise).  We found the QPOs to be strongest at high
energy (this is in accordance with all previous results on kHz QPOs),
and for that reason the 5.7--24.1 keV band provides the most
significant detection. We have used this band throughout our analysis
. The results of the fits were corrected for background and
differential dead time (see van der Klis 1989).  The errors were
determined using $\Delta\chi^2 = 1.0$ and the upper limits using
$\Delta\chi^2 = 2.71$, which corresponds to the 95~\% confidence
level. The reduced $\chi^2$ values of the fits were all $\sim 1$.

\section{Results \label{results}}

Combining all data of the August 1 observation, two simultaneous QPOs
are detected at frequencies of 900.1\pp2.4 Hz and 1176.2\pp2.9 Hz
(Fig. \ref{powerspectrum}, left). The rms amplitude (5.7--24.1 keV,
see \S \ref{observations}) and FWHM were 3.4\pp0.5 \% and 15\pp6 Hz,
and 4.5\pp0.5 \% and 27\pp10 Hz, respectively.  During the first data
segment of this observation we detect only the higher frequency QPO at
1176.2\pp2.2 Hz, with a FWHM of 16\pp8 Hz and an amplitude of
4.5\pp0.6 \% rms. The 95 \% confidence upper limit on the amplitude of
a QPO near 900 Hz with a FWHM of 15 Hz is 3.1 \% rms.  During the
second segment we find two QPOs at frequencies of 898.3\pp3.3 Hz and
1158.6\pp9.0 Hz, and a FWHM and rms amplitude of 22\pp8 Hz and
5.3\pp0.7 \%, and 37\pp21 Hz and 5.2\pp1.0 \%, respectively. During
the third and last segment we could only marginally detect the QPOs at
899.8\pp4.3 Hz (1.7$\sigma$) and 1182.9\pp13.0 Hz (2.6$\sigma$). The
FWHM and rms amplitude were 13\pp8 Hz and 3.8\pp1.0 \% for the lower
frequency QPO, and 10\pp9 Hz and 3.7\pp0.7 \% for the higher frequency
QPO.

Combining all the data of the August 31 observation only the higher
frequency QPO was detected, although marginally (2.1$\sigma$), at
1197\pp10 Hz, with a FWHM of 38\pp32 Hz and a rms amplitude of
3.8\pp0.9 \%. The lower frequency QPO had an amplitude upper limit of
2.4 \% rms (FWHM of 25 Hz).

\begin{figure}[t]
\begin{center}
\begin{tabular}{c}
\psfig{figure=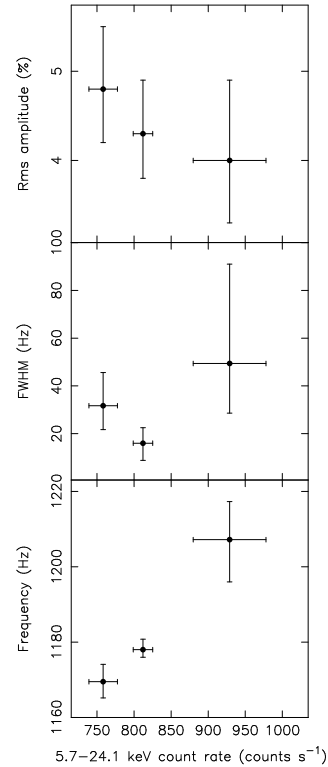,width=4.5cm}
\end{tabular}
\caption{The rms amplitude (upper), the full width at
half maximum (middle), and the frequency (lower) of the higher
frequency kilohertz QPO versus the 5.7--24.1 keV count rate. The error
bars of the count rate represent the standard deviation of the count rate
distribution in the count rate interval which was selected to
calculate the properties of the QPO.\label{countrate}}
\end{center}
\end{figure}

We combined the power spectra of the August 1 and August 31
observations to study the dependence of the kilohertz QPO properties
on 5.7--24.1 keV count rate. We selected the power spectra
corresponding to count rates below 790 counts \pers~(288 spectra),
between 790 and 840 counts \pers~(208 spectra), and above 840 counts
\pers ~(250 spectra). The selected power spectra, which were mixed in
time (in the highest count rate selection power spectra from both
observations were combined), were averaged and fitted to determine the
kilohertz QPO properties (see Tab. \ref{tab} and Fig. \ref{countrate})
. There is a clear correlation between kilohertz QPO frequency and
count rate, similar to what was observed in other burst sources.

\begin{figure}[htb]
\begin{center}
\begin{tabular}{c}
\psfig{figure=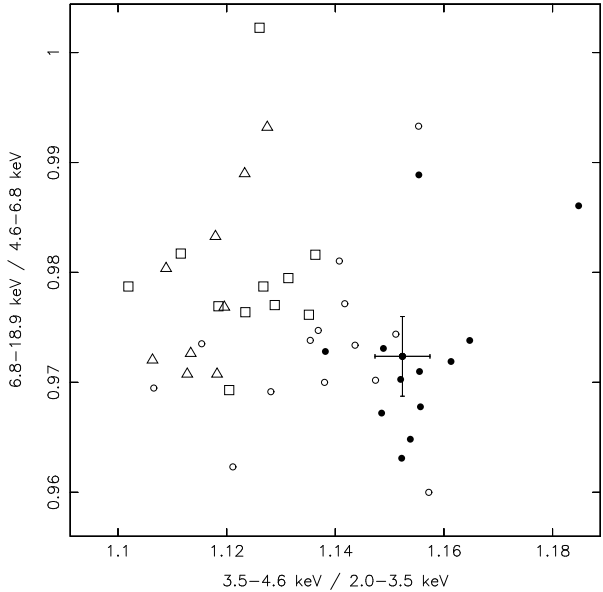,width=8cm}
\end{tabular}
\caption{The X-ray color-color diagram of the 1996 August 1
and August 31 observations. The open circles, the open triangles, and
the open squares are data segments one, two, and three, respectively,
of the August 1 observation. The filled circles are all data of the
August 31 observation. Typical statistical error bars for the colors
are shown. All points are 256-s averages. The background was not
subtracted; part of the variations in the colors may therefore be due
to background fluctuations. \label{cd}}
\end{center}
\end{figure}

We made a single combined X-ray color-color diagram (CD) for both
observations (Fig. \ref{cd}) to study the dependence of the the
kilohertz QPO properties on position of the source in the CD.  From
Fig. \ref{cd} it is clear that the kilohertz QPO properties are
correlated with the position of the source in the color-color
diagram. When the source is at the upper left part in the CD (low soft
colors, high hard colors) two kilohertz QPOs are visible in the power
spectrum. When the soft color increases and the hard color decreases
both QPOs decrease in amplitude (the lower frequency QPO becomes
undetectable) and the frequency of the higher frequency QPO increases.
The overall count rate of the source also increases in this sense
(from upper left to lower right in the CD).

Due to the low signal-to-noise ratio in the individual energy channels
the dependence of the rms amplitude of the kilohertz QPOs on photon
energy could not be determined in detail. In the energy range 2.8--4.6
keV both QPOs were not detectable, with rms amplitude upper limits of
3.4 \%. In the energy range 4.6--8.2 keV the rms amplitude of the
lower and higher frequency QPO were 4.0\pp0.7 \% and 3.9\pp0.6 \%,
respectively. In the energy range 8.2--24.1 keV the lower frequency
QPO was not detectable (amplitude upper limit of 10 \% rms); the rms
amplitude of the higher frequency QPO was 5.1\pp1.1 \%.

The VLFN during the August 1 observation had a rms amplitude of
5.9\pp0.6 \% and a power law index of 1.61\pp0.07. The peaked HFN had
a amplitude of 4.6\pp0.4 \% rms, a power law index of 0.8\pp0.6, and a
cut-off frequency of 22\pp13 Hz.  The VLFN during the August 31
observation had an rms amplitude of 7.1\pp0.6 \%, and a power law
index of 1.51\pp0.06. Although no HFN could be detected in that
observation a QPO at the frequency 26.9\pp2.3 Hz is seen in the power
spectrum (Fig. \ref{powerspectrum}, right). The FWHM and amplitude of
this QPO is 11.1\pp5.2 Hz and 3.4\pp0.6 \% (5.7--24.1 keV).

\section{Discussion \label{discussion}}

\begin{deluxetable}{ccccccc}
\tablecolumns{8}
\footnotesize
\tablewidth{0pt}
\tablecaption{The kilohertz QPOs versus 5.7--24.1 keV
count rate \label{tab}}
\tablehead{
\colhead{}&  \multicolumn{3}{c}{Lower frequency QPO}&\multicolumn{3}{c}{Higher frequency QPO} \\
Count rate$^{\rm a}$&  rms Amplitude    & FWHM  & Frequency & rms Amplitude  & FWHM & Frequency  \\
\colhead{counts \pers} &(\%)    & (Hz) & (Hz)      & (\%) & (Hz) & (Hz)  }
\startdata
758\pp19&   4.3\pp0.6 & 20\pp3  &  903.3\pp2.7 &  4.8\pp0.7     & 32\pp14 & 1169.5\pp4.6    \\
812\pp13&   $<4.7^{\rm b}$   &         &              &  4.3\pp0.6     & 16\pp7  & 1178.0\pp2.8 \\
929\pp49&   $<2.9^{\rm b}$   &         &              &4.0\pp0.9$^{\rm c}$ & 49\pp42 & 1207\pp11\\
\enddata
\tablenotetext{}{NOTE--All errors of the QPO properties correspond to
$\Delta\chi^2=1$. The upper limits correspond to a 95 \% confidence level. }
\tablenotetext{a}{Errors are the standard deviation of the count rate distribution in the count
rate interval which was selected to calculate the properties of the
QPO.}
\tablenotetext{b}{The upper limits are for QPOs near 900 Hz, with a FWHM of
50 Hz.}
\tablenotetext{c}{At a 2.1 $\sigma$ level.}
\end{deluxetable}

We detected for the first time two simultaneous kilohertz QPOs in KS
1731$-$260, at frequencies near 898 Hz and 1159 Hz. The amplitudes of
these QPOs increase with increasing photon energy.  Both QPOs are
visible at the lowest count rates. At higher count rates the lower
frequency QPO disappears, the amplitude of the higher frequency QPO
decreases, and its frequency increases. At the highest count rates the
higher frequency QPO is only marginally detected at 1207\pp11 Hz. If
that QPO is real then its frequency is the highest so far observed in
an X-ray binary. When the lower frequency QPO is detected its
frequency is consistent with being constant. The minor differences in
its frequency between slightly different data sets are within the
statistical errors.

When we combine all power spectra from the August 1 observation, the
peak separation of the two QPOs (276.1\pp3.8 Hz) is not consistent
with being equal to half the frequency of the 523.92\pp0.05 Hz
oscillations found by Smith et al. (1997) in a type I burst. However,
this inconsistency is artifical. In order to obtain a better
signal-to-noise ratio we summed the power spectra of all three data
segments obtained during this observation. During the first segment we
see only the higher frequency QPO near 1176 Hz, during the second
segment we see both QPOs near 898 and 1159 Hz, and during the third
segment only the higher frequency peak can be significantly detected
near 1183 Hz.  Combination of these three data segments gives rise to
an artifical increase of the peak separation, due to the different
focus in time of the two QPOs.  When both kilohertz QPOs are present
(segment 2) the peak separation is 260.3\pp9.6 Hz, which is almost
exactly half the oscillation frequency (261.96) during the type I
burst. This result strongly supports models in which the neutron spin
frequency is 261.96 Hz (3.8 ms), the higher frequency QPO a preferred orbital
frequency around the neutron star, and the lower frequency QPO the
beat between these two. Such a model, with the preferred orbital
radius the magnetospheric radius, was already suggested for the three
QPOs found in 4U 1728$-$34 (Strohmayer et al. 1996).  Miller, Lamb \&
Psaltis (1997) proposed a model where the preferred radius is the
sonic radius. A problem for these models is the case of Sco X-1, in
which the frequency difference is not constant (van der Klis et
al. 1997) as would be expected.

The positive correlation between soft color and X-ray count rate and
the negative correlation between hard and soft color (Fig. \ref{cd}),
the properties of the VLFN, and the decrease in the HFN strength with
count rate taken together indicate that KS 1731$-$260 probably is an
atoll source, which in our observations was in the lower banana
branch, with the inferred accretion rate somewhat higher on August 31
than on August 1. The large scatter in the CD may in part be due to
background fluctuations. The 27 Hz QPO is not uncommon in atoll
sources (Hasinger \& van der Klis 1989; Strohmayer et al. 1996) and
seems to be closely related to the HFN (``peaked HFN'').

So, the properties of the kilohertz QPOs in KS 1731$-$260 are
consistent with what is known from observations of kilohertz QPOs in
other atoll sources. The QPOs decrease in amplitude with mass
accretion rate (as inferred from the count rate, the 2-10 keV fluxes,
and the position in the CD), and where we can measure it (in the
higher frequency QPO) the frequency of the QPO increases with mass
accretion rate.

\acknowledgments

This work was supported in part by the Netherlands Organization for
Scientific Research (NWO) grant PGS 78-277 and by the Netherlands
Foundation for Research in Astronomy (ASTRON) grant 781-76-017. This
research has made use of data obtained through the High Energy
Astrophysics Science Archive Research Center Online Service, provided
by the NASA/Goddard Space Flight Center.


\begin{references}


\reference{}Berger, M., van der Klis, M., van Paradijs, J., Lewin,
W. H. G., Lamb, F., Vaughan, B., Kuulkers, E., Augusteijn, T., Zhang,
W., Marshall, F. E., Swank, J. H., Lapidus, I., Lochner, J. C.,
Strohmayer, T. E.
1996,
\apj, 469, L13

\reference{}Bradt, H. V., Rothschild, R. E., Swank, J. H.
1993,
\aaps, 97, 355


\reference{} Ford, E., Kaaret, P., Tavani, M., Barret, D., Bloser, P.,
Grindlay, J., Harmon, B. A., Paciesas, W. S., Zhang, S. N.
1996,
\apj, 475, L123

\reference{}Hasinger, G., van der Klis, M.
1989,
\aap, 225, 79

\reference{}Miller, C., Lamb, F. K., Psaltis, D.
1997,
\apj, submitted 


\reference{}Smale, A. P., Zhang, W., White, N. E.
1996,
\iaucirc, 6507

\reference{}Smith, D. A., Morgan, E. H., Bradt, H.
1997,
\apj Letters, in press

\reference{}Strohmayer, T. E., Zhang, W., Swank, J. H., Smale, A.,
Titarchuk, L., Day, C.
1996,
\apj, 469, L9

\reference{}Sunyaev, R.,
1989,
\iaucirc, 4839 

\reference{}van der Klis, M.  
1989, 
NATO ASI C262: {\it Timing Neutron Stars}, H. \"Ogelman and
E. P. J. van den Heuvel (eds.), Kluwer, p.~27.



\reference{}van der Klis, M., Swank, J. H., Zhang, W., Jahoda, K.,
Morgan, E. H., Lewin. W. H. G., Vaughan, B., van Paradijs, J.
1996a, 
\apj, 469, L1

%\reference{}van der Klis, M., van Paradijs, J., Lewin. W. H. G., Lamb,
%F. K., Vaughan, B., Kuulkers, E., Augusteijn, T.
%1996c,
%\iaucirc, 6428

\reference{} van der Klis, M., Wijnands, R. A. D., van Paradijs, J.,
Lewin. W. H. G., Lamb, F. K., Vaughan, B., Kuulkers, E., Psaltis, D.,
Dieters, S.
1996b,
\iaucirc, 6511

\reference{} van der Klis, M., Homan, J., Wijnands, R., Kuulkers, E.,
Lamb, F. K., Psaltis, D., Dieters, S., van Paradijs, J., Lewin,
W. H. G., Vaughan, B.
1997a
\iaucirc, 6565

\reference{}van der Klis, M., Wijnands, R. A. D. , Chen, W., Horne, K.
1997b,
\apj Letters, submitted


\reference{} Wijnands, R. A. D., van der Klis, M., van Paradijs, J.,
Lewin, W. H. G., Lamb, F. K., Vaughan, B., Kuulkers, E., Augusteijn,
T.
1996,
\iaucirc, 6447

\reference{} Wijnands, R. A. D., van der Klis, M., van Paradijs, J.,
Lewin, W. H. G., Lamb, F. K., Vaughan, B., Kuulkers, E.,
1997
\apj, 479, L00

\reference{}Zhang, W., Lapidus, I., White, N. E., Titarchuk, L
1996,
\apj, 469, L17

\reference{}Zhang, W., Lapidus, I., Swank, J. H., White, N. E.
1997,
\iaucirc, 6541

\end{references}
\end{document}